\documentclass{article}
\pdfoutput=1

    \usepackage[nonatbib,final]{neurips_2019}

\usepackage[utf8]{inputenc} 
\usepackage[T1]{fontenc}    
\usepackage{hyperref}       
\usepackage{url}            
\usepackage{booktabs}       
\usepackage{amsfonts}       
\usepackage{nicefrac}       
\usepackage{microtype}      

\usepackage[table,xcdraw]{xcolor}

\usepackage{tikz}
\usetikzlibrary{matrix, positioning, shapes.geometric, arrows, calc, backgrounds, decorations.pathreplacing, mindmap,trees}

\usepackage{subcaption}

\usepackage{array}
\newcolumntype{P}[1]{>{\centering\arraybackslash}p{#1}}
\usepackage{enumitem}

\usepackage{amssymb}
\usepackage{pifont}
\newcommand{\cmark}{\ding{51}}%
\newcommand{\xmark}{\ding{55}}%

\definecolor{pastelblue}{RGB}{81,96,145}
\definecolor{pastellblue}{RGB}{116,190,193}
\definecolor{pastellgreen}{RGB}{173,235,190}
\definecolor{pastellyellow}{RGB}{238,243,173}
\definecolor{pastellorange}{RGB}{243,213,173}

\usepackage{filecontents,pgfplots}

\usepackage{mathtools}
\usepackage{bm}
\usepackage{esvect}

\usepackage{lipsum}

\tikzset{%
	do path picture/.style={%
		path picture={%
			\pgfpointdiff{\pgfpointanchor{path picture bounding box}{south west}}%
			{\pgfpointanchor{path picture bounding box}{north east}}%
			\pgfgetlastxy\x\y%
			\tikzset{x=\x/2,y=\y/2}%
			#1
		}
	},
	sin wave/.style={do path picture={    
			\draw [line cap=round] (-3/4,0)
			sin (-3/8,1/2) cos (0,0) sin (3/8,-1/2) cos (3/4,0);
	}},
	cross/.style={do path picture={    
			\draw [line cap=round] (-2/4,-2/4) -- (2/4,2/4) (-2/4,2/4) -- (2/4,-2/4);
	}},
	plus/.style={do path picture={    
			\draw [line cap=round] (-3/4,0) -- (3/4,0) (0,-3/4) -- (0,3/4);
	}}
}

\tikzset{jump/.style={
	to path={
		let \p1=(\tikztostart),\p2=(\tikztotarget),\n1={atan2(\y2-\y1,\x2-\x1)} in
		(\tikztostart) -- ($($(\tikztostart)!#1!(\tikztotarget)$)!0.15cm!(\tikztostart)$)
		arc[start angle=\n1+180,end angle=\n1,radius=0.1cm] -- (\tikztotarget)}
},
jump/.default={0.5}
}

\title{Co-Attentive Cross-Modal Deep Learning for Medical Evidence Synthesis and Decision Making}

\author{%
  Devin Taylor, Simeon Spasov and Pietro Li\`o\\ 
  Department of Computer Science and Technology, University of Cambridge\\
  \texttt{\{dt475, ses88, pl219\}@cam.ac.uk} \\
}

\pgfplotsset{compat=1.14}

\begin{document}

\maketitle

\vspace{-10pt}

\begin{abstract}

Modern medicine requires \textit{generalised} approaches to the synthesis and integration of multimodal data, often at different biological scales, that can be applied to a variety of evidence structures, such as complex disease analyses and epidemiological models. However, current methods are either slow and expensive, or ineffective due to the inability to model the complex relationships between data modes which differ in scale and format. We address these issues by proposing a cross-modal deep learning architecture and \textit{co-attention mechanism} to accurately model the relationships between the different data modes. Differentiating Parkinson's Disease (PD) patients from healthy patients forms the basis of the evaluation. The model outperforms the previously published state-of-the-art unimodal analysis by $2.35~\%$, while also being $53~\%$ more parameter efficient than the industry standard cross-modal model. Furthermore, the evaluation of the attention coefficients allows for \textit{qualitative insights} to be obtained. Through the coupling with bioinformatics, a novel link between the interferon-gamma-mediated pathway, DNA methylation and PD was identified. We believe that our approach has general applications and could optimise the process of medical evidence synthesis and decision making in an actionable way.

\end{abstract}

\section{Introduction}
\label{sec:intro}

One of the biggest challenges facing the medical field is the ability to understand and diagnose complex diseases. Complex diseases are caused by a combination of genetic, environmental and lifestyle factors \cite{craig2008complex}. Understanding the influence of each of these factors requires the study of biomedical data which can exist in drastically different biological scales and formats. Complex disease analyses primarily focus on unimodal data and do not necessarily consider the interdependence between the different modes of data available. This work aims to exploit these relationships through the use of Machine Learning (ML) to diagnose, and gain data-driven insights into, complex diseases.

Cross-modal deep learning is a machine learning technique that utilises information obtained from multiple different data sources, which all independently encode information about a specific outcome, to obtain a greater understanding of the problem domain. Cross-modal models commonly use specialist subnetworks to independently extract features from the different data modes, before concatenating the features together to perform a final prediction. This approach has been successfully used in audio-visual speech classification \cite{ngiam2011multimodal}, object recognition \cite{eitel2015multimodal}, and cancer subtype classification \cite{liang2015integrative}. However, the shortfall of this approach is understanding the complex relationships between data modes. This is due to the dependence on a final feed-forward neural network to model all relationships between the independently extracted features in the concatenated latent space.

This work leverages off of cutting-edge research in Neural Machine Translation (NMT) and Visual Question Answering (VQA) to propose a cross-modal model which focuses on better modelling these complex relationships. Specifically, the model adapts the concepts of co-attention from VQA~\cite{lu2016hierarchical} to multi-head self-attention from NMT \cite{vaswani2017attention} to present the Multi-Head Co-Attention (MHCA) model. Attention has previously shown benefits in cross-modal learning for graph structured data \cite{deac2019attentive}. The architecture proposed in this research is a generalised solution which places an emphasis on learning a \textit{joint similarity distribution} between the data modes, irrespective of the biological scales and format of the input biomedical data.


Observing the attention weights further enables qualitative validation of the model predictions. Interpretability is a dominant theme in ML for health \cite{cleophas2015machine, cabitza2017unintended}. Existing solutions primarily focus on independent modules which run in parallel to unimodal models monitoring the model properties, producing statistically-based interpretations \cite{ribeiro2016should, luo2016automatically, lahav2018interpretable}. Whereas, the solution presented in this work has the interpretability as an integral component of the model architecture.


Parkinson's Disease (PD) is a complex neurodegenerative disorder which primarily affects the motor functionality of a patient \cite{lang1998parkinson}. The cause of the disease is not yet well understood, resulting in ineffective preventative and treatment measures. This has seen the incident and death rate of PD increase year-on-year \cite{dorsey2018global}. Consequently, PD is one of the most active areas of medical research. There have been very few applications of ML in PD research. These applications either focus on patient diagnosis \cite{salvatore2014machine} or symptom and biomarker identification \cite{kubota2016machine, tsanas2012accurate}. All of these applications have only considered unimodal data, no focus has been placed on leveraging the diverse multimodal data available. Therefore, PD is used as a case study for the remainder of this paper.

\section{Dataset and Preprocessing}
\label{sec:data}

This study makes use of the brain SPECT images and corresponding DNA methylation (DNA-m) data from the PPMI database\footnote{Data can be accessed at \url{https://www.ppmi-info.org/}, last accessed 15 August 2019.}. SPECT images are 3-D images of the brain which are used to model dopamine-transporter functionality, while DNA-m data is the only epigenetic marker for which a detailed mechanism of mitotic inheritance has been described \cite{bird2002dna}. This provides a lifetime record of an individual's environmental exposures. The dataset consists of $121$ healthy and $287$ PD patients.

The SPECT images have the shape $91 \times 109 \times 91$ and were co-registered and normalised to the range $[0, 1]$. The DNA-m data has $765,373$ $\beta$-value features, filtered according to \cite{zhou2017comprehensive}, in the range $[0, 1]$. $\beta$-values are a ratio of intensities between methylated and unmethylated cytosine-guanine dinucleotide (CpG) sites. 

Feature selection, using the recursive feature selection algorithm with an XGBoost model, was used to address the curse of dimensionality associated with the DNA-m data (see Appendix~\ref{app:rfe} for details). This resulted in 441 highly predictive features. Upfront feature selection, as opposed to using a neural network to generate a compressed latent representation of the original input, enabled interpretability for the DNA-m data by allowing for a one-to-one mapping between attention weights and features.


\section{Multi-Head Co-Attention Model Design}

\begin{figure}[htbp]
	\centering
		\begin{tikzpicture}[scale=.75, every node/.style={scale=.75}]
		\tikzstyle{sqr} = [rectangle, rounded corners, minimum width=1cm, text width=1.5cm, minimum height=.75cm,text centered, draw=black, line width=.3mm]
		
		\node (m1) [text width=.3cm] {$\vv{{s}}$};
		\node (m2) [text width=.35cm, below of=m1, yshift=-1cm] {$\vv{{m}}$};

		\node (enc1) [sqr, right of=m1, xshift=.5cm] {encoder};
		\node (enc2) [sqr, right of=m2, xshift=.5cm] {encoder};

		\node (mhca) [sqr, right of=enc1, xshift=2cm, yshift=-1cm, minimum height=1.5cm] {MHCA};

		\node (mult1) [circle, draw, cross, right of=mhca, xshift=.75cm, yshift=.5cm] {};
		\node (mult2) [circle, draw, cross, right of=mhca, xshift=.75cm, yshift=-.5cm] {};

		\node (p1) [circle, draw, plus, right of=mult1, xshift=.75cm] {};
		\node (p2) [circle, draw, plus, right of=mult2, xshift=-.cm] {};

		\node (p3) [circle, draw, plus, right of=p1, xshift=-.25cm, yshift=-.5cm] {};

		\node (l1) [sqr, right of=p3, xshift=.75cm] {linear};
		\node (out) [text width=.25cm, right of=l1, xshift=.75cm] {};

		\draw[->, line width=.3mm] (m1) -- (enc1);
		\draw[->, line width=.3mm] (m2) -- (enc2);
		\draw[->, line width=.3mm] (enc1) -| +(1.5, -.5) |- ($(mhca.west)+(0, .25)$);
		\draw[->, line width=.3mm] (enc2) -| +(1.5, .5) |- ($(mhca.west)+(0, -.25)$);
		\draw[->, line width=.3mm] ($(mhca.east)+(0, .5)$) -- node [yshift=.25cm] {${\alpha_q}$} (mult1);
		\draw[->, line width=.3mm] ($(mhca.east)+(0, -.5)$) -- node [yshift=.25cm] {${\alpha_v}$} (mult2);

		\draw[->, line width=.3mm] (mult1) -- node [yshift=.3cm, xshift=-.28cm] {$\vv{{q'}}$} (p1);
		\draw[->, line width=.3mm] (mult2) -- node [yshift=.3cm, xshift=-.cm] {$\vv{{v'}}$} (p2);

		\draw[->, line width=.3mm] (p1) -| (p3);
		\draw[->, line width=.3mm] (p2) -| (p3);
		\draw[->, line width=.3mm] (p3) -- node [yshift=.25cm, xshift=.0cm] {$\vv{{h}}$} (l1);
		\draw[->, line width=.3mm] (l1) -- node [yshift=.0cm, xshift=1.25cm] {prediction} (out);

		\node at (5.5, -2) {$\vv{{v}} (=\vv{{k}})$};

		\draw[->, line width=.3mm] ($(enc1.east)+(0, .0)$) -| +(.62, .3) -| node [yshift=.0cm, xshift=-3.55cm] {$\vv{{q}}$} (mult1);
		\draw[->, line width=.3mm] ($(enc2.east)+(0, -.0)$)  -| +(.62, -.3) -| node [yshift=.55cm, xshift=-3.55cm] {$\vv{k}$} (mult2);
		\draw[->, line width=.3mm] ($(enc1.east)+(0, .0)$) -| +(.62, .6) -| ($(p2.north)+(0, 1)$) to[jump=0.185] (p2);
		\draw[->, line width=.3mm] ($(enc2.east)+(0, -.0)$) -| +(.62, -.6)  -| ($(p1.south)+(0, -1)$) to[jump=0.185] (p1);
		
		\end{tikzpicture}
		\caption{Overview of the proposed MHCA model for two data modes, see text for symbol definitions. Note, the proposed model is generalisable to different data modes.}
		\label{fig:overview}
\end{figure}
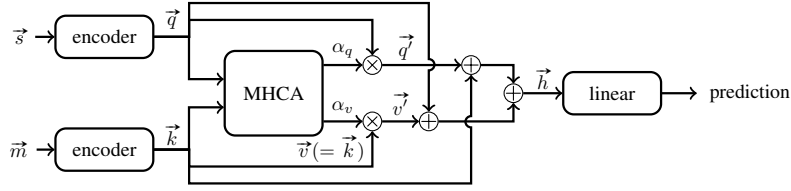

Figure~\ref{fig:overview} provides an overview of the MHCA model applied to the SPECT images, $\vv{{s}}$, and DNA-m data, $\vv{{m}}$. Specialist encoder networks generate the keys, $\vv{{k}}$, queries, $\vv{{q}}$, and values, $\vv{{v}}$, for the attention mechanism, where the keys and values are the same. The MHCA mechanism generates the attention coefficients, ${\alpha_q}$ and ${\alpha_v}$, from the input keys and queries. These, coefficients are used to weight the queries and values, producing $\vv{{q'}}$ and $\vv{{v'}}$, respectively. Residual connections are added to each of the weighted outputs to account for information loss with the attention mechanism \cite{he2016deep}. The weighted outputs are then added together to generate the hidden representation $\vv{{h}}$, a concept adapted from VQA for co-attention mechanisms \cite{lu2016hierarchical}, and passed through a pointwise linear layer with sigmoid activation to perform the final prediction. This highlights how the different data modes influence the features extracted from one another instead of the features being extracted independently\footnote{Code available at: \url{https://github.com/Devin-Taylor/multi-head-co-attention}}.

\subsection{Encoders}
\label{sec:encoders}

The encoder subnetworks extract important and correlated features between data modes and manipulate the output to be of the format \texttt{features} $\times$ \texttt{embedding}. The DNA-m encoder simply assigns an embedding dimension of $64$ to the input as the feature selection is done upfront.


The SPECT encoder consists of $3$ blocks of a $3 \times 3 \times 3$ convolution layer followed by a $2 \times 2 \times 2$ max pooling layer with ReLU activation \cite{nair2010rectified}. Each convolution layer consisted of $16$, $32$, and $64$ filters, respectively. This was followed by a convolution layer with $64$ filters and no activation function as it is responsible for generating the SPECT features. The final output is of dimension $7 \times 9 \times 7 \times 64$, which is collapsed into the format $441 \times 64$. The fact that the convolution and max pooling operators maintain the spatial relationships between their inputs and outputs makes it possible to upsample the attention weights back to the dimensions of the original inputs for interpretability.

\subsection{Multi-Head Co-Attention Mechanism}

The MHCA mechanism leverages the scaled dot-product attention from the Transformer network \cite{vaswani2017attention}. Equation~\ref{eqn:sdp} defines the similarity score between two features, $e_{ij}$, as a function of the dot product between the SPECT features $\text{\textbf{q}}=\{\vv{q_1}, \vv{q_2}, \cdots, \vv{q_n}\}$ and the DNA-m features $\text{\textbf{k}}=\{\vv{k_1}, \vv{k_2}, \cdots, \vv{k_n}\}$, and the feature dimensionality of the DNA-m, $d_k$. Where $T$ refers to the transpose operator.

\vspace{-5pt}

\begin{equation}
e_{ij} = \dfrac{\vv{q_i} \vv{k_j}^T}{\sqrt{d_k}}
\label{eqn:sdp}
\end{equation}


The Transformer network \cite{vaswani2017attention} determines a single vector of attention coefficients from the similarity matrix, as translation is directed. Given that cross-modal learning is not directed, this research leverages the commutative property of the dot product operator to determine two vectors of attention coefficients. This exploits the fact that the distribution of attention weights will likely be different when looking from the keys to queries compared to the queries to keys. The attention coefficients for the values ($\text{\textbf{v}}=\text{\textbf{k}}=\{\vv{v_1}, \vv{v_2}, \cdots, \vv{v_n}\}$) and queries are defined in Equations~\ref{eq:alphak} and \ref{eq:alphaq}, respectively. The hidden space, $\vv{{h}}$, is obtained by taking the sum of the weighted data modes, $\vv{{q'}}$ and $\vv{{v'}}$, and residual connections as defined in Equation~\ref{eq:add}. LayerNorm is the layer normalisation operator \cite{ba2016layer}.

\noindent\begin{minipage}{.5\linewidth}
\begin{equation}
\alpha_{v_{ij}} = \dfrac{\text{exp}(e_{ij})}{\sum_{m=1}^{n}\text{exp}(e_{im})}
\label{eq:alphak}
\end{equation}
\end{minipage}%
\begin{minipage}{.5\linewidth}
\begin{equation}
\alpha_{q_{ij}} = \dfrac{\text{exp}(e_{ij}^T)}{\sum_{m=1}^{n}\text{exp}(e_{im}^T)}
\label{eq:alphaq}
\end{equation}
\end{minipage}

\begin{equation}
\vv{h_{i}} = \text{LayerNorm}\left[\left(\vv{k_i} + \sum_{j=1}^{n} \alpha_{q_{ij}}\vv{q_{j}}\right) + \left(\vv{q_i} + \sum_{j=1}^{n} \alpha_{v_{ij}}\vv{v_{j}}\right)\right]
\label{eq:add}
\end{equation}

\section{Experiments and Results}


Experimentation is centred around classifying healthy patients from PD patients. The model is compared to the unimodal XGBoost DNA-m model (see Appendix~\ref{app:rfe}), the unimodal SPECT encoder network with a final linear layer (\textit{SpectNet}), a cross-modal model which concatenates independently extracted features together (\textit{ConcatNet}), and the previously published State-Of-The-Art (SOTA) 3-D CNN for PPMI PD classification \cite{choi2017refining}. The dataset was split into 80~\% train and 20~\% holdout test sets. Three runs of 5-fold cross validation (CV) were performed for each experiment, where each fold is a distinct split in the dataset. $10~\%$ data augmentation was used for all experiments\footnote{Augmentation achieved using \texttt{multiaug} library: \url{https://github.com/Devin-Taylor/MultiAug}}. A batch size of $16$, Adam optimisation \cite{kingma2014adam} with a learning rate of $3\times10^{-5}$, and $200$ epochs were used. MHCA input features were projected into four linear sub-spaces (heads). Table~\ref{tab:ppmi_results} presents the accuracy, AUROC, and parameter count for each of the experiments. The results show the MHCA model improving on the SOTA results by $2.35~\%$, while also requiring $53~\%$ fewer parameters than the \textit{ConcatNet} model. These results highlight the benefits of learning a joint similarity distribution in cross-modal learning.


\begin{table}[htbp]
	\centering
	\footnotesize
	\caption{Accuracy, AUROC, and parameter count test set results. Mean and standard deviation for 3 runs of 5-fold CV presented. 3D-CNN results taken from literature due to lack of reproducibility.}
	\label{tab:ppmi_results}
	\begin{tabular}{|P{.18\linewidth}|P{.075\linewidth}|P{.075\linewidth}|P{.14\linewidth}|P{.14\linewidth}|P{.17\linewidth}|}
		\hline
		\textbf{Model} & \textbf{DNA-m} & \textbf{SPECT} & \textbf{Accuracy (\%)} & \textbf{AUROC} & \textbf{Parameter count} \\ \hline
		\rowcolor[HTML]{EFEFEF}
		\textbf{XGBoost} & \cmark & \xmark & $95.08 \pm 0.00$  & $0.917 \pm 0.00$ & $-$ \\ \hline
		\textbf{SpectNet} & \xmark & \cmark & $88.07 \pm 2.54$ & $0.879 \pm 0.01$ & $3,793,378$ \\ \hline
		\textbf{3D-CNN from \cite{choi2017refining}} & \xmark & \cmark & $96.00$ & $-$ & $-$ \\ \hline
		\rowcolor[HTML]{EFEFEF}
		\textbf{ConcatNet} & \cmark & \cmark & $97.12 \pm 0.58$ & $0.968 \pm 0.00$ & $3,824,066$ \\ \hline
		\rowcolor[HTML]{EFEFEF}
		\textbf{MHCA} & \cmark & \cmark & $\textbf{98.35 $\pm$ 0.58}$ & $\textbf{0.976 $\pm$ 0.00}$ & $\textbf{2,020,883}$ \\ \hline
	\end{tabular}
\end{table}

\begin{figure}[htbp]
	\centering
	\hspace{2pt}
	\begin{subfigure}[t]{0.34\columnwidth}
		\centering
    	\scalebox{.85}{
    		\includegraphics[width=\columnwidth]{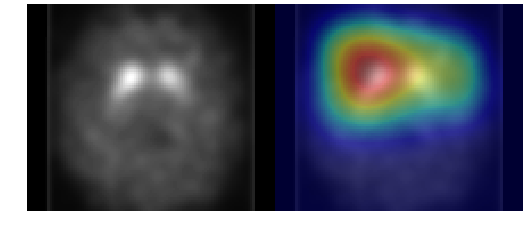}}
    	\caption{PD, top, $37/91$.}
    	\label{fig:pd_2_top}
    	\scalebox{.85}{
    		\includegraphics[width=\columnwidth]{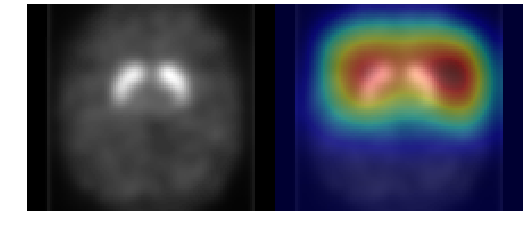}}
    	\caption{Healthy, top, $35/91$.}
    	\label{fig:heathly_1_top}
	\end{subfigure}
	\begin{subfigure}[t]{0.29\columnwidth}
	\centering
	\scalebox{.85}{
		\includegraphics[width=\columnwidth]{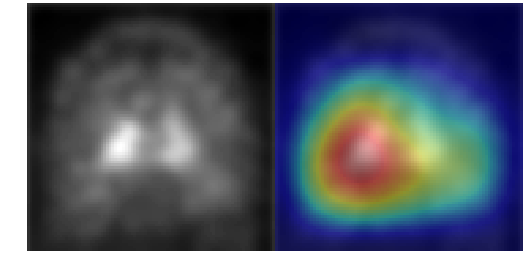}}
	\caption{PD, back, $62/109$.}
	\label{fig:pd_2_back}	
	\scalebox{.85}{
		\includegraphics[width=\columnwidth]{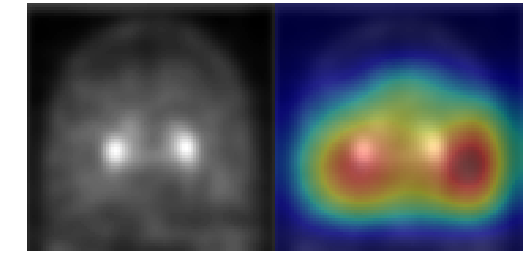}}
	\caption{Healthy, back, $60/109$.}
	\label{fig:healthy_1_back}
\end{subfigure}
\begin{subfigure}[t]{0.25\columnwidth}
	\centering
	\scalebox{.85}{
		\includegraphics[width=\columnwidth]{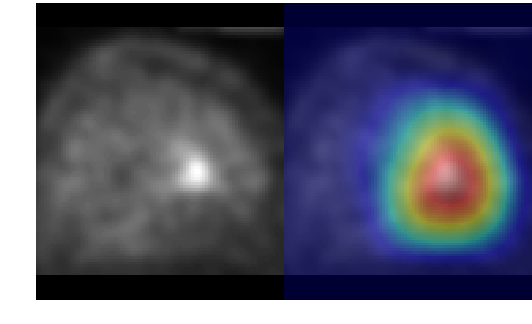}}
	\caption{PD, side, $50/91$.}
	\label{fig:pd_2_side}
	\scalebox{.85}{
		\includegraphics[width=\columnwidth]{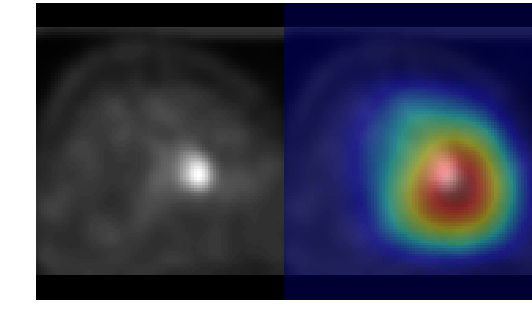}}
	\caption{Healthy, side, $52/91$.}
	\label{fig:healthy_1_side}
\end{subfigure}
	\scalebox{.85}{
		\begin{subfigure}[t]{0.01\columnwidth}
			\vspace*{-2.2cm}\includegraphics[height=2.05cm]{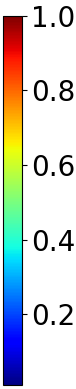}
			\par\medskip
			\vspace*{.55cm}\includegraphics[height=2.05cm]{cbar}
	\end{subfigure}}
	\caption{Patient samples with weight mask overlay. Top: PD patient, bottom: healthy patient. The subcaptions are of the format: label, angle, volume slice number. Figures best observed in colour.}
	\label{fig:patients}
\end{figure}

Qualitatively, Figure~\ref{fig:patients} provides slices from the different angles of the brain for a healthy and PD patient. Notably, the weight masks validate existing research \cite{booth2015role} by highlighting that the primary differentiating factor in SPECT images is the dopaminergic cell loss that is associated with PD patients. This is evident as the model focuses equally on both putamens (white regions) in the healthy patient but only on the putamen without the apparent degradation in the PD patient. 

PD is commonly linked to environmental factors, hence DNA-m plays a fundamental role in understanding the gene-environment interactions \cite{miranda2017implications}. To investigate these interactions the CpG sites that the attention weights placed the most emphasis on were selected (136 sites). The \texttt{Bioconductor} R package \cite{hansen2016bioconductor} was used to obtain the set of 90 genes corresponding to the sites. \texttt{DAVID}, a bioinformatics tool, was used to determine the biological pathways relevant to the genes \cite{huang2008bioinformatics, huang2009systematic}. This analysis identified the interferon-gamma-mediated (IFN-$\gamma$) pathway, alternative splicing, and transcriptional activator activity as the most important pathways. IFN-$\gamma$, with the highest p-value ($2.6\times10^{-3}$), has featured in recent studies linking gene expression data and the PD phenotype \cite{mount2007involvement, barcia2011ifn, liscovitch2014differential}. These studies all focus on gene expression patterns, whereas this study presents the novel finding that links IFN-$\gamma$ to PD through DNA-m data. This finding is important as DNA-m patterns are considered more reliable and stable than gene expression patterns. It also suggests that changes in DNA-m, as a result of lifestyle, could influence the progression of PD. These results highlight how the interpretability can be used to not only support patient diagnosis but also facilitate exploratory analyses of data.

\section{Conclusion}

A generalised cross-modal deep learning model for multimodal data with different biological scales is presented. The model uses a novel co-attention mechanism to learn a joint similarity distribution between data modes, thus better modelling the complex relationships between them. PD forms the basis of the evaluation, exceeding the SOTA classification results by $2.35~\%$, while being $53~\%$ more parameter efficient. The model produced granular and informative interpretations, linking novel biomarkers to the IFN-$\gamma$ pathway. The results obtained highlight the value of attention in cross-modal learning. Future work will focus on scalability and memory efficiency for higher-dimensional data, and studying how perturbations to one data mode influence another data mode.

\clearpage
\newpage



\small
\bibliographystyle{vancouver}
\bibliography{ms.bib}

\clearpage
\newpage

\appendix

\renewcommand\thetable{\thesection.\arabic{table}}
\setcounter{table}{0}

\section{DNA-Methylation Preprocessing}
\label{app:rfe}

This appendix provides additional information on the upfront feature selection process performed for the DNA-methylation data, which is necessary for reproducing the results obtained in the main text.

Feature selection is a powerful machine learning technique which aims to reduce the number of features in a dataset by selecting a subset of predictive features. Feature selection involves training a model, which has built-in feature importance, to obtain a ranking of the features in terms of their predictive power. Thereafter, the top $n$ features are selected based on a user-defined metric.

An XGBoost model is used in this study. XGBoost is a popular gradient boosting algorithm with built-in ensembling \cite{Chen:2016:XST:2939672.2939785}. The model is popular for the ability to handle data with few samples, making it suitable for the given application. The XGBoost model was combined with the Recursive Feature Elimination (RFE) algorithm to perform the final feature selection. RFE is a recursive algorithm that trains an XGBoost model on the set of available features. RFE then uses the feature importance from the XGBoost algorithm to remove a subset of least predictive features. It repeats this process until a user-defined number of features are selected. RFE is significantly more powerful than simply using the feature importance from a single XGBoost model as it accounts for the fact that the importance of features might change in the presence or absence of other features. The model was tasked on classifying Parkinson's Disease patients from healthy patients using the PPMI dataset. Table~\ref{tab:xgb_hyperparams} defines the hyperparameters for the models, obtained performing a parameter sweep. The number of features to be selected by the RFE algorithm was set to 441 to ensure the dimensionality with the SPECT images is consistent.

\begin{table}[ht]
	\centering
	\footnotesize
	\caption{Hyperparameters for RFE XGBoost feature selection model.}
	\label{tab:xgb_hyperparams}
	\begin{tabular}{|l|c|c|}
		\hline
		\textbf{Parameter} & \textbf{Algorithm} & \textbf{Value} \\ \hline
		\textbf{learning\_rate} & XGBoost & 0.1 \\ \hline
		\textbf{n\_estimators} & XGBoost & 100 \\ \hline
		\textbf{max\_depth} & XGBoost & 6 \\ \hline
		\textbf{min\_child\_weight} & XGBoost & 2 \\ \hline
		\textbf{gamma} & XGBoost & 0 \\ \hline
		\textbf{subsample} & XGBoost & 0.8 \\ \hline
		\textbf{colsample\_bytree} & XGBoost & 0.8 \\ \hline
		\textbf{nthreads} & XGBoost & 15 \\ \hline
		\textbf{scale\_pos\_weight} & XGBoost & 1 \\ \hline
		\textbf{n\_features\_to\_select} & RFE & 441 \\ \hline
		\textbf{step} & RFE & 0.05 \\ \hline
	\end{tabular}
\end{table}

In order to validate that reducing the set of features did not negatively impact the data's predictive power, the XGBoost model was trained on the original dataset and on the reduced dataset. The dataset was split into 80~\% train and 20~\% holdout test sets. The results obtained can be seen in Table~\ref{tab:xgboost_results}. Table~\ref{tab:xgboost_results} shows that the reduced set of features actually improved the results. This can be attributed to the unfiltered data suffering from the curse of dimensionality. The reduced model forms a baseline for future comparisons.

\begin{table}[htbp]
	\footnotesize
	\centering
	\caption{Mean and standard deviation test set results for the XGBoost model before and after feature selection using the RFE algorithm. Results are from 3 runs of 5-fold cross validation.}
	\label{tab:xgboost_results}
	\begin{tabular}{|P{.15\linewidth}|P{.2\linewidth}|P{.2\linewidth}|}
		\hline
		\textbf{Number of features} & \textbf{Accuracy (\%)} & \textbf{AUROC} \\ \hline
		\textbf{765,373} & $67.21 \pm 0.00$ & $0.477 \pm 0.00$ \\ \hline
		\textbf{441} & $\textbf{95.08 $\pm$ 0.00}$ & $\textbf{0.917 $\pm$ 0.00}$ \\ \hline
	\end{tabular}
\end{table}

\end{document}